\documentclass[conference]{IEEEtran}
\usepackage{cite}
\usepackage{amsmath,amssymb,amsfonts}
\usepackage{algorithmic}
\usepackage{graphicx}
\usepackage{textcomp}
\usepackage{makecell}
\usepackage{listings}
\usepackage{hhline}
\usepackage{xcolor}
\usepackage{url}

\begin{document}
\title{Accelerating advection for atmospheric modelling on Xilinx and Intel FPGAs}

% author names and affiliations
% use a multiple column layout for up to three different
% affiliations
\author{\IEEEauthorblockN{Nick Brown}
\IEEEauthorblockA{EPCC at the University of Edinburgh\\
The Bayes Centre,\\
47 Potterrow, \\
Edinburgh, UK\\
Email: n.brown@epcc.ed.ac.uk}}

% make the title area
\maketitle

\begin{abstract}
Reconfigurable architectures, such as FPGAs, enable the execution of code at the electronics level, avoiding the assumptions imposed by the general purpose black-box micro-architectures of CPUs and GPUs. Such tailored execution can result in increased performance and power efficiency, and as the HPC community moves towards exascale an important question is the role such hardware technologies can play in future supercomputers.

In this paper we explore the porting of the PW advection kernel, an important code component used in a variety of atmospheric simulations and accounting for around 40\% of the runtime of the popular Met Office NERC Cloud model (MONC). Building upon previous work which ported this kernel to an older generation of Xilinx FPGA, we target latest generation Xilinx Alveo U280 and Intel Stratix 10 FPGAs. Exploring the development of a dataflow design which is performance portable between vendors, we then describe implementation differences between the tool chains and compare kernel performance between FPGA hardware. This is followed by a more general performance comparison, scaling up the number of kernels on the Xilinx Alveo and Intel Stratix 10, against a 24 core Xeon Platinum Cascade Lake CPU and NVIDIA Tesla V100 GPU. When overlapping the transfer of data to and from the boards with compute, the FPGA solutions considerably outperform the CPU and, whilst falling short of the GPU in terms of performance, demonstrate power usage benefits, with the Alveo being especially power efficient. The result of this work is a comparison and set of design techniques that apply both to this specific atmospheric advection kernel on Xilinx and Intel FPGAs, and that are also of interest more widely when looking to accelerate HPC codes on a variety of reconfigurable architectures.
\end{abstract}

\IEEEpeerreviewmaketitle

\section{Introduction}
Demand for more accurate, more ambitious scientific and engineering simulations continues to place increasing pressure on modern HPC. A key question raised by exascale is, in the coming years, what mix of hardware is most appropriate for future supercomputers. One such technology, popular in other fields but yet to gain widespread adoption for HPC is that of Field Programmable Gate Arrays (FPGAs) which provide a large number of configurable logic blocks sitting within a sea of configurable interconnect. Whilst limited hardware capabilities and esoteric programming environments historically limited their update by scientific software developers, recent years have seen very significant advances made by the vendors. This covers both the hardware capability and also the programming environments for FPGAs, now meaning that this technology is more accessible to HPC developers than ever before. Whilst such advances mean the physical act of programming FPGAs is now becoming more a question of software development rather than hardware design, there are still many fundamental questions to answer around the types of workloads that FPGAs can accelerate, the most appropriate dataflow algorithmic techniques, and which specific FPGA hardware is most suited to which workloads. 

% Rephrase this (And previous) as very similar to previous FPGA paper!
The Met Office NERC Cloud model (MONC) \cite{easc} is an open source, high resolution, atmospheric modelling framework that employs Large Eddy Simulation (LES) to study the physics of turbulent flows and further develop and test physical parametrisations and assumptions used in numerical weather and climate prediction. This is used by scientists for activities including the development and testing of the Met Office Unified Model (UM) boundary layer scheme \cite{lock1998}, convection scheme \cite{petch2001} and cloud microphysics \cite{hill2014}. Advection, computing the movement of quantities through the air due to kinetic effects, is responsible for a significant amount of the overall model runtime, approximately 40\%, with one of the main schemes being that of the Met Office's Piacsek and Williams (PW) \cite{pwadvection}. It is therefore interesting to explore the role of novel hardware in accelerating this kernel, as this not only impacts the MONC model and its established user base, but furthermore techniques and lessons learnt can be applied to other HPC workloads too.

% Three main features - drata trnas to/from and compute
In previous work \cite{advection_fpga} \cite{brown2019s} we ported the Met Office PW advection scheme to an ADM-PCIE-8K5 board, which contains a Xilinx Kintex UltraScale KU115-2 FPGA, using High Level Synthesis (HLS). Whilst this work demonstrated promise, performing comparably to an 18-core Broadwell CPU, both the FPGA hardware and software environment imposed a limit on the performance that was delivered. Modern data centre FPGAs, the Xilinx Alveo and Intel Stratix 10 families, are a generation beyond the KU115-2, which itself is not really designed for high performance workloads. Furthermore, the previous work was conducted at a time prior to the release of Xilinx's Vitis programming environment, meaning that whilst we were able to write the kernel in C++ and synthesise to the underlying Hardware Description Language (HDL) via HLS, there were a number of additional hardware oriented tasks required such as manually developing the shell via Vivado block design. Since then there have been significant developments in the field of FPGAs, with the release of Xilinx's Vitis framework, the wide spread availability of the Xilinx Alveo range of FPGA boards, and growing maturity of Intel Stratix 10 and the Quartus Prime Pro ecosystem. Therefore it is instructive to revisit this kernel, from the perspective of developing a design which is portable across the latest Xilinx and Intel FPGA data-centre hardware via these more advanced tool chains. This enables us to not only more accurately gauge the benefits that FPGAs can deliver in accelerating this specific atmospheric advection scheme compared to latest CPU and GPU, but furthermore enables a comparison between vendors to understand the portability of the underlying dataflow algorithmic design and most appropriate techniques.

% HLS

In this paper we explore the redesign of this advection kernel with the aim of FPGA vendor portability and performance. The paper is organised as follows; In Section \ref{sec:background} we provide a technical overview of the kernel as well as describing previous work in this area, and the hardware configurations utilised throughout the rest of the paper. This is then followed by Section \ref{sec:kernel_design} which focuses on developing a portable dataflow design for FPGAs along with the specific implementation details for both Xilinx and Intel technologies. This section also provides a performance comparison of a single HLS kernel on Intel and Xilinx FPGAs, with the CPU and GPU for context. Section \ref{sec:multi-kernel} then expands the performance discussion by scaling up the number of FPGA kernels, and comparing not only the raw kernel performance but also considering the overhead of data transfer to and from the device via PCIe. This section compares the performance, power, and power efficiency for our kernel against that of a 24-core Cascade Lake Xeon Platinum CPU and NVIDIA Tesla V100 GPU. The paper then draws to a close in Section \ref{sec:conclusions} where we discuss conclusions and identify further work based upon the vendor technology road-maps.

\section{Background and Related Work}
\label{sec:background}

High Level Synthesis (HLS) has empowered programmers to write C or C++ code for FPGAs and for this to be translated into the underlying Hardware Description Language (HDL). Not only does HLS avoid the need to write code at the HDL level for FPGAs, significantly improving productivity and opening up the programming of FPGAs to a much wider community, but furthermore the tooling developed by vendors also automates the process of integration with the wider on-chip infrastructure such as memory controllers and interconnects as well as making emulation and performance profiling more convenient. With both the Xilinx Vitis and Intel Quartus Prime Pro tool chains the programmer writes their host code in OpenCL, which is the technology also used for writing device side code for Intel FPGAs in Quartus Prime Pro. However, whilst device side OpenCL is supported by Xilinx's Vitis framework, it is more common to leverage their bespoke C++ HLS technology which involves the use of pragma style hints to drive the tooling. These are the configurations used in this paper, OpenCL on the host for both vendors, with OpenCL on the device for Intel and HLS C++ on the device for Xilinx.

However, HLS is not a silver bullet and whilst the physical act of programming FPGAs has become much easier, one must still recast their Von-Neumann style algorithms into a dataflow style \cite{koch2016fpga} in order to obtain best performance. There are numerous previous and on-going activities investigating the role that FPGAs can play in accelerating HPC codes, such as \cite{brown2020exploring}, \cite{lfricfpga}, \cite{yang2019fully}, and whilst there have been some successes, many struggle to perform well against the latest CPUs and GPUs. An important question is not only what are the most appropriate dataflow techniques, but also whether both Xilinx and Intel FPGA technologies can be effectively programmed from a single dataflow algorithmic design. Of course the actual code will look somewhat different, with different syntax and potentially different pragmas, but fundamentally for a kernel such as our advection scheme, are there the same general set of concerns between the vendors that the programmer must address for good performance?

\subsection{The existing MONC advection FPGA kernel}
% ALL double precision!

Advection is the movement of grid values through the atmosphere due to wind. At around 40\% of the overall runtime, advection is the single longest running piece of functionality in the MONC model where the code loops over three fields;  \emph{U}, \emph{V} and \emph{W}, representing wind velocity in the \emph{x}, \emph{y} and \emph{z} dimensions respectively. With the coordinate system oriented with \emph{z} in the vertical, \emph{y} in the horizontal, and \emph{x} in the diagonal, this scheme is called each timestep of the model and calculates advection results, otherwise known as \emph{source terms}, of the three fields. Processing each grid cell involves 63 double precision operations, and this advection kernel is a stencil based code, of depth one, requiring neighbouring values across all the three dimensions per grid cell. Listing \ref{lst:pw_orig} illustrates a sketch of the Fortran code for calculating the \emph{U} field source in a single timestep term.

\begin{lstlisting}[frame=lines,caption={Illustration of the PW advection scheme for the u field only},label={lst:pw_orig}, numbers=left]
do i=1, x_size
  do j=1, y_size
    do k=2, z_size
      su(k, j, i) = tcx * (u(k,j,i-1) * (u(k,j,i) + u(k,j,i-1)) - u(k,j,i+1) * (u(k,j,i) + u(k,j,i+1)))
      
      su(k, j, i) = su(k, j, i) + tcy * (u(k,j-1,i) * (v(k,j-1,i) + v(k,j-1,i+1)) - u(k,j+1,i) * (v(k,j,i) + v(k,j,i+1)))
      
      if (k .lt. z_size) then
        su(k, j, i) = su(k, j, i) + tzc1(k) * u(k-1,j,i) * (w(k-1,j,i) + w(k-1,j,i+1)) - tzc2(k) * u(k+1,j,i) * (w(k,j,i) + w(k,j,i+1))
      else 
        su(k, j, i) = su(k, j, i) + tzc1(k) * u(k-1,j,i) * (w(k-1,j,i) + w(k-1,j,i+1))
      end if
    end do
  end do
end do
\end{lstlisting}

\begin{figure}[h]
\centering
\includegraphics[scale=0.55]{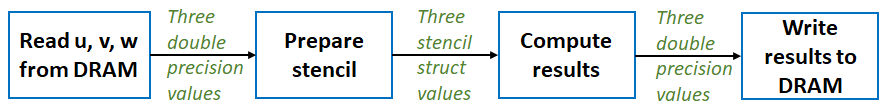}
\caption{Illustration of the dataflow design of the existing kernel from \cite{brown2019s}}
\label{fig:dataflow_existing_design}
\end{figure}

In \cite{advection_fpga} and \cite{brown2019s} where we ported the kernel to an ADM-PCIE-8K5 board using HLS we first converted the Fortran code to C++, then leveraging the \emph{dataflow} HLS pragma to run four components, \emph{loading data}, \emph{stencil preparation}, \emph{computation of results}, and \emph{writing results} concurrently with data streaming between these. This is illustrated in Figure \ref{fig:dataflow_existing_design}, and whilst it resulted in reasonable kernel level performance, the code was very complicated, especially around the preparation of the stencil where all required values were packaged together into a single structure. Moreover, the overhead of data transfer to and from the board via PCIe was found to be very significant, and motivated by CUDA streams we adopted a bespoke data chunking and streaming approach to overlap data transfer with computation where possible. However, the specialised nature of the Alpha Data host-side API meant that this was complex, brittle, and lacked generality. 

\subsection{Hardware setup}

For the runs contained in this paper we use a Xilinx Alveo U280 and Intel Stratix 10. The Alveo U280 contains an FPGA chip with 1.08 million LUTs, 4.5MB of on-chip BRAM, 30MB of on-chip URAM, and 9024 DSP slices. This PCIe card also contains 8GB of High Bandwidth Memory (HBM) and 32GB of DDR DRAM on the board. Codes for the Alveo are built with Xilinx Vitis framework version 2020.2.

The Stratix 10 is mounted on a Bittware 520N card containing a Stratix 10 GX 2800. This contains 933,120 Adaptive Logic Modules (ALMs), 1.87MB of on-chip MLAB memory, 28.6MB of M20K memory, and 5760 DSP blocks. The card also contains 32GB of DDR DRAM. Codes for the Stratix 10 are built with Intel's Quartus Prime Pro version 20.4. Host code is built using GCC version 8.3.

Throughout the experiments detailed in this paper we compare against a 24-core Xeon Platinum (Cascade Lake) 8260M CPU, and NVIDIA Tesla V100 GPU.

\section{Kernel design and implementation}
\label{sec:kernel_design}

\begin{figure}[h]
\centering
\includegraphics[scale=0.50]{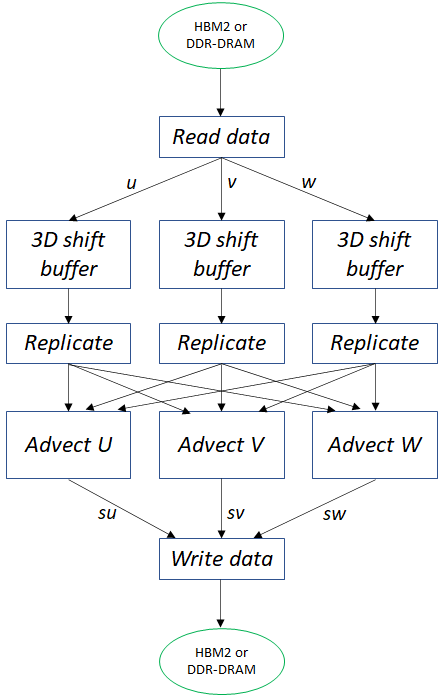}
\caption{Illustration of the dataflow redesign of our advection kernel}
\label{fig:dataflow_design}
\end{figure}

Our objective for a single kernel is to calculate the advection source terms for fields \emph{U}, \emph{V}, \emph{W} of a grid cell each clock cycle. The previous dataflow design of this kernel, as illustrated in Figure \ref{fig:dataflow_existing_design} resulted in complex code, and we therefore redesigned this, adopting a more generalised approach. Our new dataflow design is illustrated in Figure \ref{fig:dataflow_design}, where each box is an independent region that is running concurrently and streaming data to other regions. After input data has been read in this is then passed to the \emph{shift buffer} stage which implements a 3D shift buffer to provide the stencil values needed for each individual grid cell calculation. Advection, calculating each field's source terms, requires all fields as input values, hence the \emph{replicate} stages in Figure \ref{fig:dataflow_design} which replicates the input stream values for each field advection. The advection stages perform the double precision floating point calculations before streaming out result values to be written back to external memory. The design methodology here is to view Figure \ref{fig:dataflow_design} as a dataflow machine which is application specific, where irrespective of how it is implemented on the FPGA, to achieve best performance the programmer should emphasise each stage running concurrently (e.g. not blocking on previous stages), and continually streaming result data (e.g. each cycle a new result is generated and streamed to the next stage).

As was illustrated in Listing \ref{lst:pw_orig}, in calculating source terms the kernel requires neighbouring grid cell values. Typically only 8 unique values of the 27 point 3D stencil are required for each field advection, and to achieve the two dataflow aims of no stalling and continued streaming of results, it is important to avoid duplicate external memory reads. In previous work we achieved this via a bespoke caching mechanism which only stored the stencil values required and forwarded them on. Whilst this increased code complexity significantly, it maximised usage of the more limited resources of the Kintex KU115-2 FPGA. Instead, in this work we adopted the design of a general purpose 3D shift buffer. Whilst this does increase overall resource usage, for instance we are storing and forwarding all 27 stencil points to the advection kernels, some of which are unused, this considerably simplifies the code and we felt was best to ensure portability between vendor technologies, at the cost of increasing overall resource usage.

\begin{figure}[h]
\centering
\includegraphics[scale=0.40]{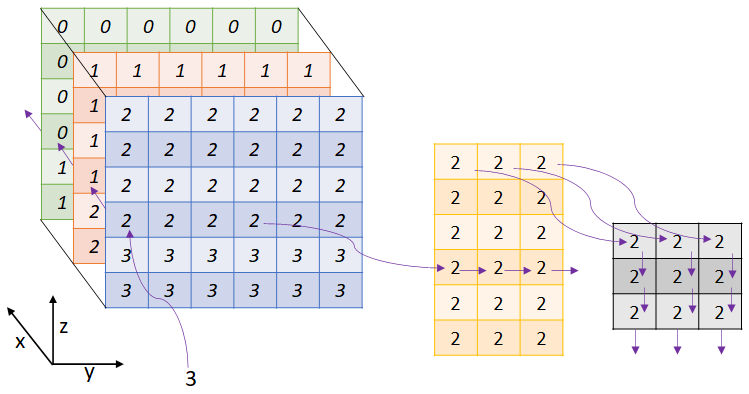}
\caption{Illustration of the 3D shift buffer design, with main 3D array containing three slices of domain in Y and Z dimension, and each slice has associated two 2D arrays (only showing the first slices's arrays for clarity)}
\label{fig:shiftbuffer_design}
\end{figure}

An illustration of our new 3D shift buffer approach is provided by Figure \ref{fig:shiftbuffer_design}, where there are three key data structures in use. Firstly, a three dimensional array (left most in Figure \ref{fig:shiftbuffer_design}) which is of size X equals 3, Y is the grid size in Y, and Z is the grid size in Z. Working upwards in a column (dimension Z), effectively this shifts in the X dimension, where a new data element is streamed in each cycle from the \emph{read data} stage and replaces the corresponding value in the current grid cell in the top X dimension. This is the blue face in Figure \ref{fig:shiftbuffer_design}, where the value \emph{3} is replacing the value \emph{2}, and this replaced value then replaces the corresponding value in the next slice (orange colour in Figure \ref{fig:shiftbuffer_design})  which itself replaces the value in the third slice (green colour). 

Each slice of the 3D array in Figure \ref{fig:shiftbuffer_design} has an associated 2D rectangular array of size Z being the grid size in Z and Y being 3. This can be imagined as sliding across, in the Y dimension, its corresponding slice of the 3D array. This 2D array is effectively shifting the data in the Y dimension, where a value is read from the large 3D array slice and replaces the value in the left most current line of the rectangular 2D array, and the line itself shifts one place to the right. Lastly, each 2D array has a three by three array associated with it, each cycle values from the 2D rectangular array are loaded into the corresponding columns and these shifted down.  

In Figure \ref{fig:shiftbuffer_design} a single example of the 2D arrays are provided for the first (blue) slice, this is for clarity of presentation and in reality these are array exist for each slice of the 3D buffer. Therefore there are three 2D rectangular arrays (in yellow) and three 3x3 boxes (in grey), one for each slice of the main 3D array. Each cycle the three, 3x3 arrays for this field in question are streamed to the advection stage, representing the required 27-point stencil. 

Therefore, once filled, this 3D shift buffer means that per clock cycle one input value is consumed and a complete 27-point stencil generated for the advection stage. Furthermore, given correct partitioning, there are never more than two memory accesses per cycle on the 3D and 2D rectangular array which is compatible with being able to handle these accesses within a single cycle for dual ported on-chip memory. The small 3x3 arrays are implemented by both Vitis and Quartus Prime Pro as registers and-so do not require partitioning.

%For the 3D array, per cycle each slice has a read and write on a single memory element (to implement the shift) and a read on another element (to copy the value to the 2D rectangle being rolled across the 3D slice). Therefore this needs to be partitioned with the outer dimension of three only. Each yellow 2D rectangle is similar, again partitioned in one dimension (the Y) dimension of three. The last 3x3 arrays are implemented by both Vitis and Quartus Prime Pro as registers and-so do not require explicit consideration around maximum number of memory accesses and data partitioning.

\begin{figure}[h]
\centering
\includegraphics[scale=0.60]{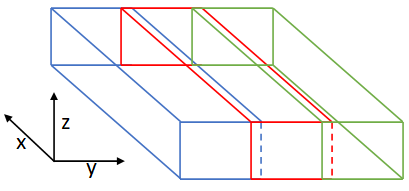}
\caption{Illustration of chunking with overlap in the Y dimension. Also illustrates the dimension system, with Z in the vertical, Y in the horizontal, and X in the diagonal}
\label{fig:chunking}
\end{figure}

There are three of these shift buffers, one for each field, and as such there is a resource limitation in terms of the amount of on-chip memory being used hold the 3D and 2D arrays. The memory required is bounded by the Y and Z dimensions only, and therefore to decouple the grid size configuration from the FPGA resources required we adopted a chunking approach where the Y dimension is split up into chunks, as illustrated by Figure \ref{fig:chunking}. The code handles each 3D chunk, before moving onto the next chunk in the Y dimension. The dotted line on the front face illustrates that, due to the 1-depth halo there is an overlap of two grid points in the Y dimension, one for the right halo of the left chunk and the other for the left halo of the right chunk.

In this manner the \emph{read data} and \emph{write data} stages load data on a chunk by chunk basis. Whilst this enables the processing of an unlimited domain size, the potential disadvantage is that it results in smaller contiguous external memory accesses, of chunk face size Y by Z, before having to move one grid point in the X dimension for the next chunk's Y-Z face. There is an overhead in accessing external data in a non-contiguous fashion \cite{vitis_bestpractice}, however in reality apart from a very small chunk size of 8 or below, this has negligible performance impact.

Another advantage of considering our design from the view point of a dataflow machine is to enable convenient reasoning about the overall theoretical performance. Each advection stage usually contains twenty one floating point operations. Given an initiation interval of one, our design means that per cycle there are usually 63 floating point operations that can run concurrently (but for the column top grid cell this reduces to 55 operations). Multiplying the clock frequency by this number provides a theoretical best performance in terms of Floating Point Operations per Second (FLOPS) that kernels should be looking to achieve. For instance, with a clock frequency of 300MHz, which is the default on the Alveo, and a column height of 64 grid cells (the default for the MONC model), the design illustrated in Figure \ref{fig:dataflow_design} can theoretically deliver a best performance of 18.86 GFLOPS. Quantifying how far kernels fall short of this figure can determine how much more opportunity there is for further kernel level optimisation.

Table \ref{fig:kernel_comparison} illustrates kernel-only performance, ignoring the overhead of data transfer to or from the card via PCIe, of our advection kernel with a medium problem size of 16 million grid cells. We compare the performance of the Xilinx Alveo U280 and Intel Stratix 10 against a 24-core Xeon Platinum Cascade Lake CPU and NVIDIA Tesla V100 GPU. The major difference between the two FPGA kernels is that the Intel kernel is running at 398MHz, whereas the Xilinx kernel is running at 300MHz. Therefore the theoretical best performance of the Intel kernel is 25.02 GFLOPS, and 18.86 GFLOPS for the Xilinx kernel. It can be seen from Table \ref{fig:kernel_comparison} that, broadly, both kernels are reaching a similar level of actual performance against their theoretical best. 

The increased clock frequency provided by the Intel means that a single advection kernel on the Stratix 10 outperforms the 24 core Xeon Platinum Cascade Lake CPU, whereas the lower frequency of the Xilinx means that it falls slightly short. Comparing against previous work in \cite{brown2019s}, running this kernel on a Xilinx Kintex UltraScale KU115-2, with eight kernels we only achieved 18.8 GFLOPS. The fact that a single kernel on the Alveo U280 achieves around 77\% of this, and the Stratix 10 out performs it by approximately 10\%, demonstrates the significant advances that the latest hardware, new software tooling, and redesign of our kernel have delivered. 

The GPU implementation tested in Table \ref{fig:kernel_comparison} is from \cite{brown2015directive}, written in OpenACC, and compiled with the Portland Group Compiler version 20.9. Whilst the performance of the GPU is impressive, it should be highlighted that this is exploiting the entirety of the GPU, whereas in Table \ref{fig:kernel_comparison} we are focusing on single FPGA HLS kernels, with a considerable amount of the chip unoccupied. Therefore whilst the GPU is a useful comparison point, the experiments detailed in Section \ref{sec:multi-kernel}, comparing against multiple FPGA kernels and considering the overhead of data transfer provide a more complete comparison.

\begin{table}[h]

%\hspace*{-0.3in}
%\begin{center}
 \centering
\begin{tabular}{ | c c c c | }
\hline
\textbf{Description} & \makecell{\textbf{Performance} \\ \textbf{(GFLOPS)}} & \makecell{\textbf{\% theoretical} \\ \textbf{performance}} & \makecell{\textbf{\% CPU} \\ \textbf{performance}} \\ \hline
1 core of Xeon CPU & 2.09 & - & - \\
24 core Xeon CPU & 15.2 & - & - \\
NVIDIA V100 GPU & 367.2 & - & 2400\% \\
Xilinx Alveo U280 & 14.50 & 77\% & 95\% \\
Intel Stratix 10 & 20.8 & 83\% & 137\% \\
\hline
\end{tabular}
%\end{center}
\caption{Kernel-only performance between Xeon Platinum CPU, NVIDIA Tesla V100 GPU, and a single FPGA kernel on the Xilinx Alveo U280 and Intel Stratix 10 for a problem size of 16 million grid points}
\label{fig:kernel_comparison}
\end{table}

%\begin{table}[h]

%\hspace*{-0.3in}
%\begin{center}
% \centering
%\begin{tabular}{ | c c c c c | }
%\hline
%\textbf{Vendor} & \textbf{LUT} & \textbf{DSP} & \textbf{BRAM} & \textbf{Build time}\\ %\hline
%Xilinx & 19\% & 4\% & 16\% & 2h 57m\\
%%Intel & 18\% & 2\% & 21\% & 3h 45m\\
%\hline
%\end{tabular}
%\end{center}
%\caption{Resource usage for a chunk size of Y=256.}
%\label{fig:kernel_utilisation}
%\end{table}

\subsection{Xilinx specific implementation details}
Each box of Figure \ref{fig:dataflow_design} is implemented via a distinct function called from within an HLS \emph{dataflow} region, connected via HLS streams. Following best practice \cite{vitis_bestpractice} we pack external data accesses to be 512 bits wide, and by default favour the high bandwidth memory (HBM2) on the Alveo U280, connecting our kernel data ports across all the HBM2 banks. At 8GB the HBM2 is large enough to hold all but our two largest grid size configurations considered in Section \ref{sec:multi-kernel}, for which we must switch to using the 32GB DDR-DRAM instead. Table \ref{fig:u280_hbm_vs_ddr} illustrates kernel only performance (i.e. ignoring the overhead of data transfer to or from the board) when using the on-chip HBM2 memory against the slower on-board DDR-DRAM memory for a variety of grid cell problem sizes. It can be seen that the high bandwidth memory delivers a significant performance benefit throughout and as such should be the preferred space when the data fits.

%We had also initially split out the reading and writing dataflow stages into separate stages, one for each input or output variable. However Vitis imposes a maximum of 32 ports across all kernels which becomes an issue as we scale up in Section X. Therefore we decided to reduce the number of ports, one for input data, one for output data, and one for constants. Vitis imposes the restriction that no two dataflow stages can concurrently access the same port, hence having a single read and single write stage. This however raised another issue around lack of bursting. Following best practice \cite{vitis_bestpractice}, we pack our memory accesses into eight double precision numbers, resulting in 512 bit wide accesses to optimise external memory efficiency. However when profiling the kernel Vitis Analyser reported that, which this was effective, the kernel was not bursting data accesses even though the appropriate pragmas had been supplied on the port definitions. Given the reasonable performance against the theoretical best we were somewhat skeptical that it was causing performance issues, but was most likely due to the fact that we had a single loop accessing three distinct memory locations, \emph{u}, \emph{v}, and \emph{w} or \emph{su}, \emph{sv}, \emph{sw} respectively. This was refactored to read into an explicit burst buffer, which improved the metrics reported by Vitis Analyser but did not increase performance.

\begin{table}[h]

%\hspace*{-0.3in}
%\begin{center}
 \centering
\begin{tabular}{ | c c c c | }
\hline
\textbf{Grid points} & \makecell{\textbf{HBM2} \\ \textbf{performance} \\ \textbf{(GFLOPS)}} & \makecell{\textbf{DDR-DRAM} \\ \textbf{performance} \\ \textbf{(GFLOPS)}} & \makecell{\textbf{DDR-DRAM} \\ \textbf{overhead}}\\ \hline
1M & 12.98 & 8.98 & 45\%\\
4M & 14.94 & 10.21 & 46\%\\
16M & 14.52 & 10.43 & 39\%\\
67M & 14.68 & 10.55 & 39\%\\
\hline
\end{tabular}
%\end{center}
\caption{Performance comparison between using HBM2 and DDR-DRAM on the Alveo U280}
\label{fig:u280_hbm_vs_ddr}
\end{table}

Internal to our HLS kernel, the arrays constituting the shift buffer are partitioned using the HLS \emph{array\_partition} pragma and reside in on-chip BRAM. Whilst we experimented with utilising the larger on-chip UltraRAM (URAM) for these instead, the increased latency of URAM meant that there was a dependency between an index's read and write operation because the write to memory must be completed in a single cycle. Consequently using URAM imposed an access latency of two cycles at 300MHz which increased the initiation interval to two of our loop. Only processing a new iteration every other cycle, effectively meaning that data would stream out every other cycle, would half the performance and as such we considered it unacceptable.

\subsection{Intel specific implementation}
Implementing our dataflow design in OpenCL is arguably less convenient than Xilinx HLS due to having to make each stage of Figure \ref{fig:dataflow_design} an explicit OpenCL kernel connected by OpenCL channels. These channels are an Intel specific extension, and all the kernels are launched from the host. This increases the verbosity of the code, but is offset by automatic optimisations such as the tooling transparently optimising external memory accesses which reduces code complexity. 

A challenge was in ensuring that the 3D shift buffer had been implemented effectively, with the tooling reporting via the HTML report that there was a memory dependency issue limiting the initiation interval (II). Whilst we initially looked to solve this by banking the memory via the \emph{bankwidth} and \emph{numbanks} qualifiers from the Intel best practice guide \cite{intel_bestpractice}, this did not solve the memory dependency issue. However it was only on the smaller, dimension of three, of the arrays illustrated in Figure \ref{fig:shiftbuffer_design} that we needed to split the arrays apart in order to avoid dependencies on the dual-ported memory. Therefore we did this manually in the code, which was not ideal but solved the issue, reducing the II to one. This array manipulation was the major specialisation required for the Intel implementation of this kernel, with the actual computation component being unchanged between the two vendors.

\subsection{Discussion}
Based on our experiences of implementing the design for both vendors it is worth considering the differences between the tool chains and how readily programmers can switch from one to another. It is our feeling that there is somewhat of a philosophical difference between the Xilinx HLS C++ and Intel OpenCL HLS approaches, where the Intel approach is more automated, placing greater emphasis on the tooling to convert the programmer's code into efficient HDL. By contrast, the Xilinx HLS approach requires more in-depth knowledge and experience but provides more insight and configuration options for tuning. External memory accesses are an example, where not only did we not need to explicitly pack data to a specific bit-width for the Stratix 10, but furthermore the Intel tooling will select the most appropriate load-store units including bursting and prefetching. 

Conversely, the array partitioning pragmas provided by Xilinx HLS felt more convenient and powerful for partitioning the shift buffer arrays than those provided by Intel to avoid memory dependency issues. Furthermore, whilst both tool chains report the performance of each loop after synthesis, the insights provided by the Xilinx HLS tooling in the analysis pane around exactly what operations are scheduled when, and how data flows between these, tends to deliver more detailed information and aid in fully understanding the effectiveness of the applied tuning pragmas. 

When using HBM2 Xilinx Alveo U280 achieves 77\% of theoretical kernel performance, which drops to 55\% when switching to DDR-DRAM. By contrast, the Stratix 10 achieves 83\% of theoretical performance using only DDR-DRAM, therefore the Intel tooling is, through automatic optimisation, ameliorate the lower performance of this memory technology.

To summarise, the Intel tooling feels like it is aimed at the software developer, whereas the Xilinx tooling expects that the developer will have a deeper understanding of the FPGA technology and will tweak aspects to gain best performance. Importantly, we have been able to produce a single kernel dataflow design, illustrated in Figure \ref{fig:dataflow_design}, and implement this in the tooling of both vendors with relative ease. Furthermore, with both versions achieving close to their theoretical maximum performance, the implementations are efficient. We believe that this demonstrates a wider point, that whilst there are implementation differences between the vendor's tool chains, many of these are somewhat superficial, with numerous underlying fundamentals shared between them and as such performance portability is possible by adopting the correct dataflow design.

\section{Multi-kernel performance and power}
\label{sec:multi-kernel}
We focused on the design and implementation of a single well performing HLS kernel in Section \ref{sec:kernel_design}. However this only occupied around 15\% of the Alveo or Stratix 10 resources, and as such can be scaled up to multiple kernels. We were able to fit six kernels on the Xilinx Alveo U280, and five kernels on the Intel Stratix 10.

\begin{figure}[h]
\centering
\includegraphics[scale=0.38]{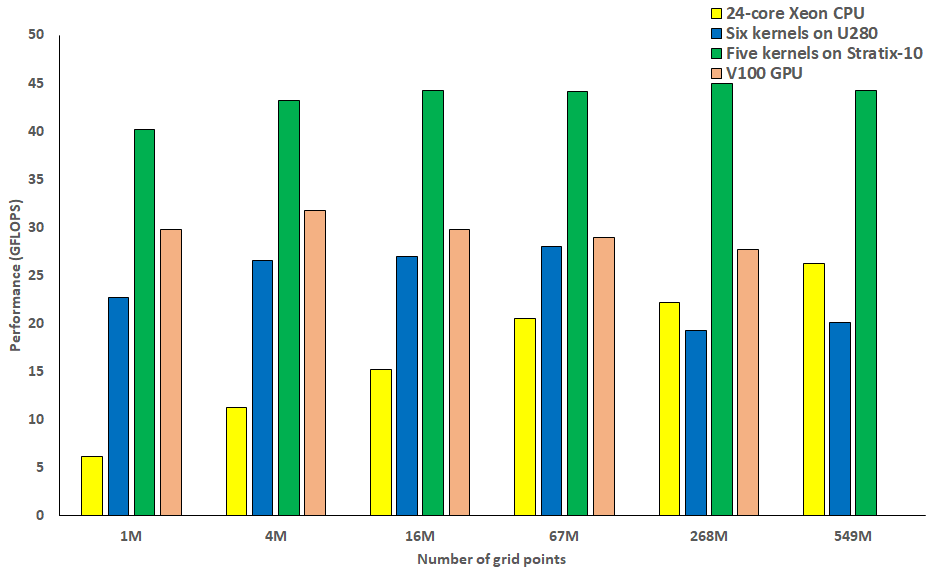}
\caption{Performance comparison between hardware technologies without any overlapping of data transfer and compute (higher is better).}
\label{fig:performance_total_no_chunking}
\end{figure}

Figure \ref{fig:performance_total_no_chunking} illustrates the performance, for a variety of grid sizes, of six FPGA kernels on the Xilinx Alveo U280 and five kernels on the Intel Stratix 10, against a 24-core Cascade Lake Xeon Platinum and V100 GPU. Reported here is the overall performance, including the overhead of data transfer to and from the FPGA or GPU boards via PCIe. It can be seen that this is significantly lower than the kernel-only results reported in Table \ref{fig:kernel_comparison}. With a problem size of 16 million grid cells, approximately 800MB of data must be transferred between the host and board via PCIe, increasing to 3.2GB, 12.8GB and 25.8GB for 67 million, 268 million, and 536 million grid cell configurations respectively. 

Clearly data transfer imposes a significant overhead on the overall performance of these accelerators, where the NVIDIA V100 GPU suffers especially compared to the raw performance delivered by the kernel itself. Furthermore there are no results for 536 million grid cells on the V100 GPU because the board contains only 16GB of memory and as such can not hold the entire 25.8GB data set required. The Intel Stratix 10 outperforms the Alveo U280 here, and this is due to a smaller overhead of data transfer to and from the card, where consistently data transfer takes approximately twice as long on the U280 than it does on the Stratix 10.

We previously encountered this same data transfer overhead issue with the ADM-PCIE-8K5 board in \cite{brown2019s}, and demonstrated that performance could be considerably improved by implementing a mechanism to overlap data transfer and kernel compute. The benefit of both the Xilinx Vitis and Intel Quartus Prime Pro tools chains is that, by writing the host code in OpenCL, then we are able to achieve such overlapping in a much more standardised, simpler manner compared to \cite{brown2019s} where we had to build upon the low-level Alpha Data bespoke API. The approach we adopted was to chunk the data up in the X dimension, where each chunk represents a smaller data-set and shorter execution of the FPGA kernel on that specific chunk. By bulk registering all data transfers to and from the FPGA, and using OpenCL events to specify that there are multiple executions of the same kernel, each with a dependency upon small data transfers, then this means that, whilst a kernel is running, input data is in-flight for subsequent kernels and result data being transferred for proceeding kernels. Therefore, with the exception of the first kernel, given a sensible chunk size then data will be present when a specific kernel starts, and with the exception of the last kernel, then the host need not block for result data transfer as these are all running concurrently.

\begin{figure}[h]
\centering
\includegraphics[scale=0.35]{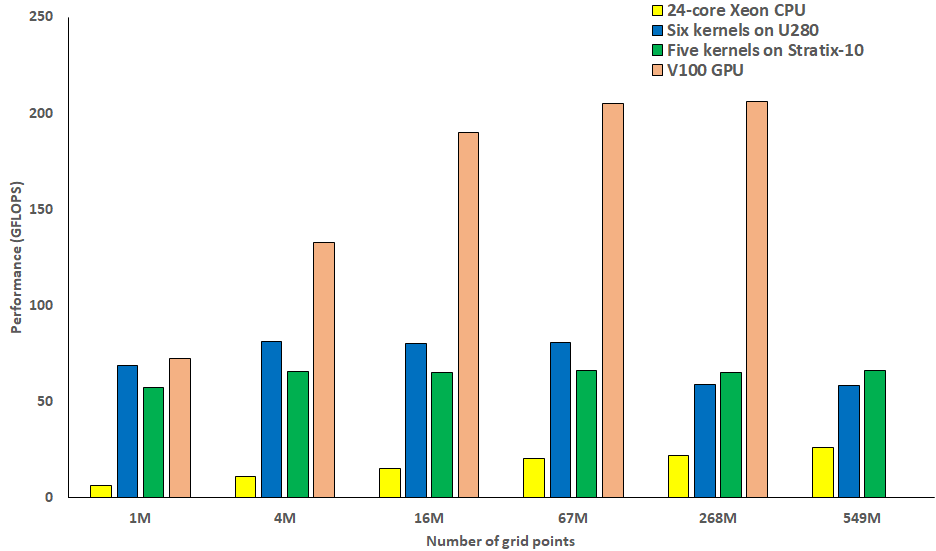}
\caption{Performance comparison between hardware technologies with overlapping of data transfer and compute (higher is better).}
\label{fig:performance_total_chunking}
\end{figure}

Performance with this overlapping of data transfer and compute is illustrated in Figure \ref{fig:performance_total_chunking}, which significantly improves the overall FPGA and GPU performance, and also makes a difference to the relative performance of the technologies. With this overlapping the V100 GPU (using CUDA streams to overlap on the GPU) outperforms all other technologies, especially at larger grid sizes. Furthermore, apart from 268 million and 536 million grid cell configurations, the Xilinx Alveo U280 outperforms the Intel Stratix 10. There are three reasons for this change in relative performance between the FPGA technologies, firstly data transfer on the Alveo is more time consuming than the Stratix 10, hence this overlapping approach benefits the Alveo the most. Secondly, we were able to fit six HLS kernels onto the Alveo U280 whereas only five onto the Intel Stratix 10, and thirdly the Alveo is able to run at 300MHz, whereas the clock frequency on the Statix-10 decreases significantly from 398MHz for a single kernel to 250MHz as the number of kernels are increased. 

In Figures \ref{fig:performance_total_chunking} and \ref{fig:performance_total_no_chunking}, it can be seen that the performance of the Alveo U280 decreases sharply at 268 million and 536 million grid cell compared with smaller domain sizes. This is because, at this configuration, the data is too large to fit in the 8GB of HBM2, and instead must use the 32GB of DDR-DRAM.

\begin{figure}[h]
\centering
\includegraphics[scale=0.35]{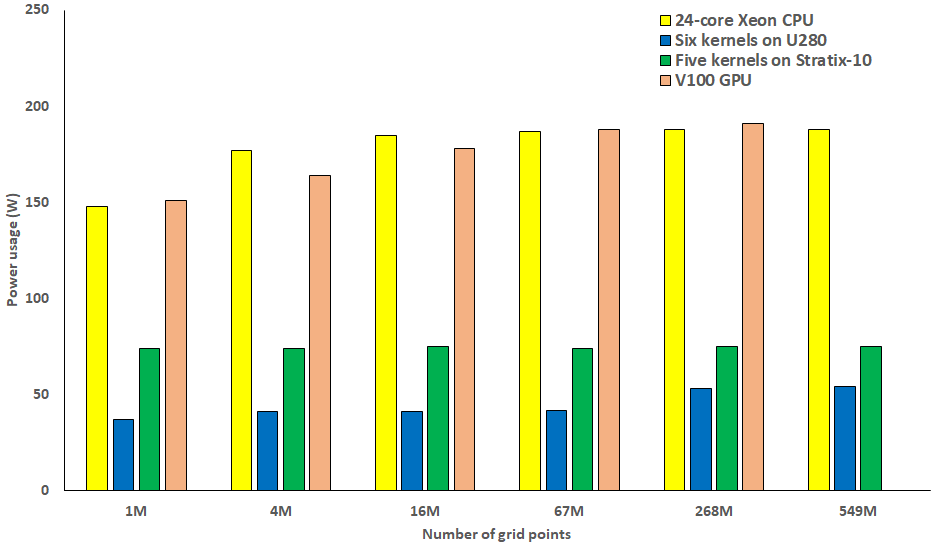}
\caption{Power usage comparison between hardware technologies with overlapping of data transfer and compute (lower is better).}
\label{fig:power}
\end{figure}

Figure \ref{fig:power} illustrates the power usage in Watts between the four different hardware technologies running our advection kernel. Power was captured on the CPU using RAPL, NVIDIA-SMI for the GPU, XRT for the Alveo, and the \emph{aocl\_mmd\_card\_info\_fn} API call on the Stratix 10. As to be expected, in absolute terms both the CPU and GPU consume significantly more power than the two FPGAs. However, there is a difference between the Stratix 10 and Alveo U280, where the Stratix 10 consumes around 50\% more power than the Alveo U280. This was unexpected, and we had initially assumed it was entirely due to the use of generally more power efficient HBM2 rather than DDR-DRAM. However reduced power draw, albeit diminished, continued at our largest grid configurations where the Alveo uses DDR-DRAM, and moving from HBM2 to DDR-DRAM saw an increase of only 12 Watts on the U280.

\begin{figure}[h]
\centering
\includegraphics[scale=0.35]{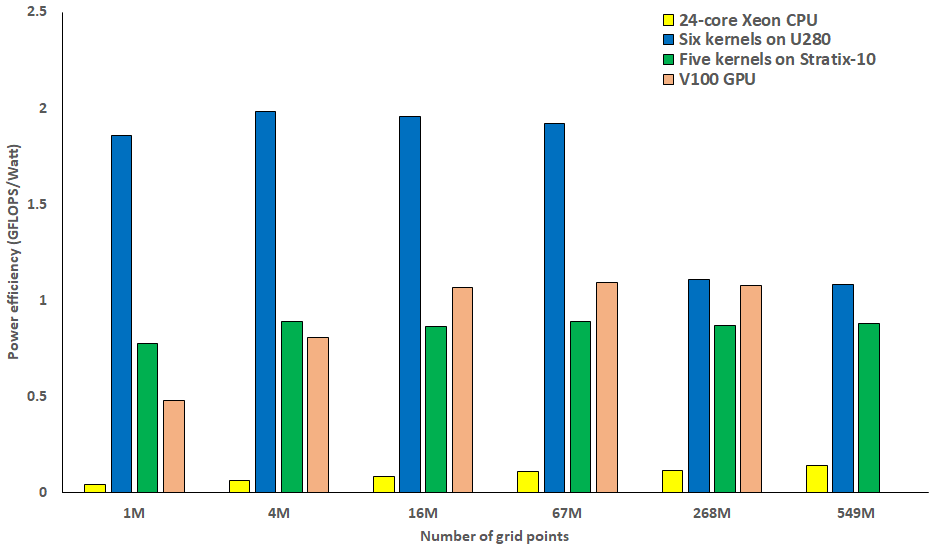}
\caption{Power efficiency comparison between hardware technologies with overlapping of data transfer and compute (higher is better).}
\label{fig:power_efficiency}
\end{figure}

Figure \ref{fig:power_efficiency} illustrates the power efficiency of our experiments, where it can be seen that the CPU's low performance and high power usage results in worst efficiency. Interesting the Stratix 10 and Alveo U280 are somewhat different here, with the increased performance of the U280 and lower absolute power draw resulting in approximately double the power efficiency of the Stratix 10 until 256 million grid points. At that point the lower performance delivered by the DDR-DRAM on the U280 results in a decrease in the power efficiency, bringing it closer to the other technologies. Whilst the GPU draws significantly more power, the higher performance delivered means that it is competitive against the Stratix 10. Whilst the Stratix 10 is more power efficient than the V100 GPU for smaller grid sizes, the V100 GPU is slightly better at larger configurations.

\section{Conclusion}
\label{sec:conclusions}
% Demonstrates the design approach too which is portable! Comment on the overall design methodology too, important!

% Hihglight theoretical performance metric too

In this paper we have explored the porting of the Piacsek and Williams (PW) atmospheric advection scheme to both Xilinx Alveo U280 and Intel Stratix 10 FPGAs. This significantly improves the performance of this kernel on a previous generation FPGA, and based upon a common dataflow design we have developed implementations for both vendors which, at the kernel level, perform close to their theoretical maximum performance. This illustrates the benefit of designing dataflow algorithms from the perspective of a dataflow machine and using the theoretical performance value as a metric to understand how well the implementation is performing. Whilst the implementation of the kernel did require some implementation specialisation between Xilinx's Vitis and Intel's Quartus Prime Pro, this is largely driven by philosophical differences between the two tool chains. Intel's Quartus Prime Pro is generally more advanced at handling automated transformations, and ultimately able to achieve a higher clock frequency for a single kernel. By contrast, Xilinx Vitis expects the programmer will be more manually involved in code optimisation, and provides to them the insight and tools required to achieve this.

When scaling to multi-kernel the performance picture between the Xilinx and Intel technologies changed, where the Intel tooling struggled more to scale the number of kernels whilst maintaining a reasonable clock frequency. For this workload, the ability to overlap data transfer and compute is crucial for performance, and we demonstrated that by adopting OpenCL on the host this is easily implementable in both tool chains and makes a significant impact on overall performance. When comparing multi-kernel performance we found that the NVIDIA Tesla V100 GPU outperforms all other hardware convincingly, with the Xilinx Alveo U280 outperforming the Intel Stratix 10 when it can leverage the high bandwidth memory, but the Stratix 10 then out performs the Alveo when DDR-DRAM must be used. As expected, the FPGAs draw considerably less power than the CPU or GPU, however the Alveo also draws around 50\% less power than the Stratix 10, meaning that the Alveo is overall most power efficient, with the Statix 10 and GPU fairly similar for that metric.

In terms of further work, exploring the role of reduced precision and fixed point arithmetic would be interesting. This could reduce the amount of resource required for our shift buffers and advection calculations, as such enabling more kernels to be fitted onto the chip. Furthermore, FPGAs with AI accelerators such as Xilinx's Versal ACAP and Intel's Stratix 10 NX look likely to dominate in the coming years and will be especially suited for accelerating lower precision arithmetic. Taking the Xilinx Versal as an example, there will be up to 400 AI engines which act as vector units clocked at around 1 GHz, each capable of performing eight single precision floating point operations per cycle. This could considerably accelerate the arithmetic component of our advection kernel, and keeping the engines fed with data will be the key, exploiting the reconfigurable fabric of the ACAP for our shift buffer design. It will also be interesting to explore how portable algorithmic techniques for exploiting these new AI accelerators are between vendors. 

We conclude that the work done in this paper demonstrates that there has been considerable advances made in the previous few years by the vendors on the hardware and software ecosystems. It is now, more than ever before, realistic to port HPC kernels to FPGAs and, if done correctly, this will convincingly surpass a modern CPU, both in terms of performance and power efficiency. Whilst the GPU is more of a challenging contender for this workload, the fact that it is possible to create performance portable dataflow designs between the two major FPGA vendors is impressive, and the next generation FPGA of technologies to be released later in 2021 will likely further close the gap between FPGAs and GPUs.

\section*{Acknowledgment}
The authors would like to thank the ExCALIBUR H\&ES FPGA testbed for access to compute resource used in this work. The authors gratefully acknowledge the computing time provided by the Paderborn Center for Parallel Computing (PC²). This work was funded under the EU EXCELLERAT CoE, grant agreement number 823691.

\bibliographystyle{IEEEtran}
\bibliography{references}

%\begin{thebibliography}{1}

%bibitem{IEEEhowto:kopka}
%H.~Kopka and P.~W. Daly, \emph{A Guide to \LaTeX}, 3rd~ed.\hskip 1em plus
%  0.5em minus 0.4em\relax Harlow, England: Addison-Wesley, 1999.

%\end{thebibliography}
\end{document}